\documentclass[aps,prl,showpacs,twocolumn,superscriptaddress,letterpaper]{revtex4}
\usepackage{amsmath}
\usepackage{amssymb}
\usepackage{amsfonts}
\usepackage{graphicx}

\def\be{\begin{equation}}       \def\ee{\end{equation}}
\def\bea{\begin{eqnarray}}      \def\eea{\end{eqnarray}}

\begin{document}
\title{Functional Renormalization Group Analysis of $\eta$-Pairing  in Iron-based Superconductors}

\author{Jing Yuan}
\affiliation{Institute of Physics, Chinese Academy of Sciences, Beijing 100190, China}
\author{Jiangping Hu}
\affiliation{Institute of Physics, Chinese Academy of Sciences, Beijing 100190,
China}
\affiliation{Department of Physics, Purdue University, West Lafayette, Indiana 47907, USA}
\affiliation{Collaborative Innovation Center of Quantum Matter, Beijing, China}

\date{\today}

\begin{abstract}
Using functional renormalization group (FRG) approach, we analyze the $\eta$ pairing in a five d-orbital model for iron-pnictides. We find that the $\eta$ pairing between two hole pockets is  a leading instability at low energy when the two hole pockets are close to a nesting condition.  The result suggests that the system can spontaneously enter into a superconducting state with both normal pairing and $\eta$ paring, and the $\eta$ pairing is driven by the inter-orbital antiferromagnetic exchange coupling  while the normal pairing is driven by  the intra-orbital  antiferromagnetic ones.  An effective model at low energy is proposed to account for the $\eta$ pairing instability.

\end{abstract}


\maketitle

In the past few years, there were rapid research developments in  the field of  iron-based superconductors. Major efforts have been devoted to understanding the pairing symmetry and pairing mechanism in these new superconductors. However  the secret of pairing symmetry and mechanism still remains elusive\cite{Hirschfeld2011,Johnston2010-review,Dagotto2013-review}.

Theoretical models for iron-based superconductors are often constructed  on an Fe-square lattice where the unit cell only includes one Fe atom.  The search of superconductivity is often limited to  $(\mathbf{k},-\mathbf{k})$ Cooper pairs (which we refer it as normal pairing)  because the pairs in an uniform superconducting state have  zero central crystal momentum\cite{BCS1957}. However, the unit cell in the true lattice structure of FeAs or FeSe layer includes two Fe atoms.  The $\mathbf{Q}=(\pi,\pi)$ wavevector in the models with a 1-Fe unit cell is a real reciprocal lattice wavevector in the real lattice with a 2-Fe unit cell. Thus,  a $(\mathbf{k},-\mathbf{k}+\mathbf{Q})$ pairing, which is named  as $\eta$-pairing\cite{Yang1989}, should also be considered as a zero central crystal momentum Cooper pair. Recently, one of us\cite{Hu2013} has pointed out that under the full lattice symmetry, the nonsymmorphic space group, $P4/nmm$, the $\eta$ pairing and the normal  pairings are classified by opposite parities. In the spin singlet pairing channel,  the  $\eta$-pairings with one-dimensional irreducible representation(IR) such as s-wave and d-wave have odd parity   and those with   two-dimensional IRs (p-wave)  have even parity.   In the past, the $\eta$ pairing channel was largely ignored.   The inter-pocket pairing in proposing the bonding-antibonding $s^\pm$\cite{Hirschfeld2011} came to the closest to resemble the $\eta$ pairing\cite{Khodas2012, Mazin2009} and a pure d-wave $\eta$ pairing was discussed in a two-orbital mode\cite{Yigao2010}.

In an effort to search for an unified state to describe both iron-pnictides and iron-chalcogenides, one of us \cite{Hu2013,Hu2014-odd,Hu2012s4} has recently suggested that the $\eta$ pairing may be significant. In fact,  to achieve  a full gapped superconducting state with a balanced sign change on Fermi surfaces\cite{Hu2014-odd,Hu2012s4}, which is required in order to consistently explain many experimental observations,  both normal pairing and $\eta$ pairing must coexist. Namely, there must be a spontaneous symmetry breaking in both $\eta$ and normal pairing channels. A possible form of the combination in the $\eta$ pairing and  normal pairing channels was suggested and some new experimental evidences have emerged\cite{Tan2013-fese,Zhang2013-impurity}. However,  the  results are mainly guided by symmetry analysis phenomenologically. It lacks of energetic understanding within a microscopic model.  In fact, the combination of the two channels that can achieve a superconducting state satisfying the requirement of the gap and sign change properties is not unique\cite{Hu2014-odd}.

In this letter, we study the $\eta$ pairing instability using the Functional Renormalization Group (FRG) method \cite{Shankar1994,Metzner2012review,Honerkamp2010,Honerkamp2001,WangZhaiLee2009,Thomale2011} on a five-band model\cite{Graser2010} of iron-pnictide with Coulomb and Hund's interactions. We find that  the $\eta$ pairing  instability between two hole pockets at $\Gamma$, $(0,0)$ and $M$, $(\pi,\pi)$ in  the 1-Fe Brillouin zone is rather strong. This $\eta$ paring instability is driven by the nesting conditions between these two hole pockets while it is known  that the normal pairing  is driven by nesting conditions between electron and hole pockets.  In the absence of hole pockets,  the $\eta$ pairing between two electron pockets and the normal pairing on electron pockets are both weak in this FRG approach with the former is slightly stronger than the latter.   These results  support the idea that the $\eta$ pairing and the normal pairing strongly coexist because they are driven by different conditions on Fermi surfaces.  Moreover, as the two hole pockets have different orbital characters, the $\eta$ pairing is mainly contributed to the inter-orbital pairing between the $d_{xy}$ and $d_{xz,yz}$ orbitals while the normal pairing is known to be dominated by intra-orbital pairings.  These results also suggest that  the previously well-known $J_1-J_2$ effective model\cite{Seo2008,Si2008} should be  modified to include the orbital dependence of the antiferromagnetic (AFM) exchange couplings. We suggest that the minimum modification is to change the nearest-neighbor  $J_1$ to be an antiferromagnetic exchange coupling between two different orbitals.

\begin{figure}
  \centering
  \includegraphics[width=5.5cm]{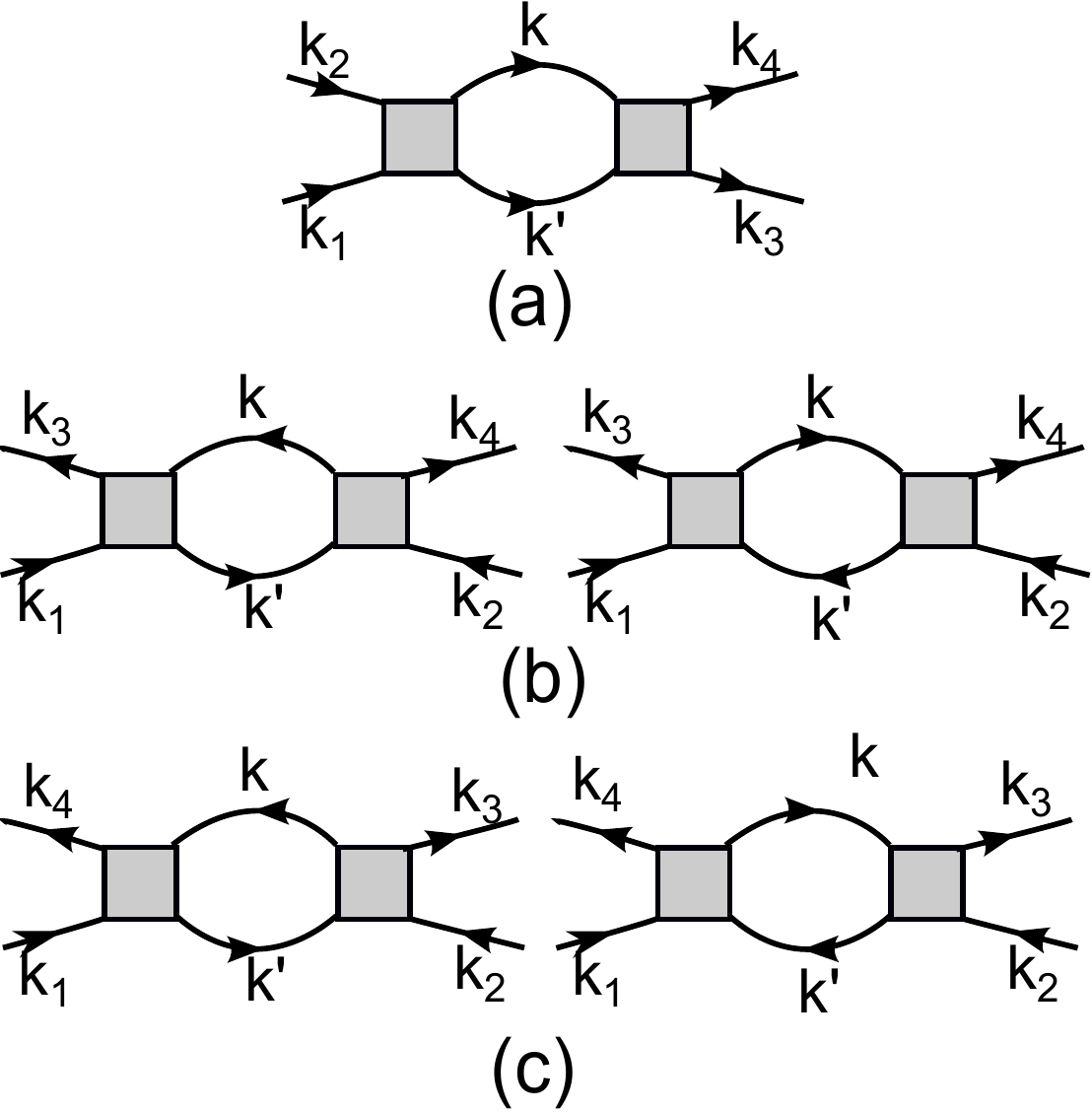}\\
  \caption{The particle-particle(a), direct particle-hole(b) and crossed particle-hole(c) channels contributing to the interaction flow.}\label{frgchannel}
\end{figure}

\begin{figure}
  \centering
  \includegraphics[width=5.5cm]{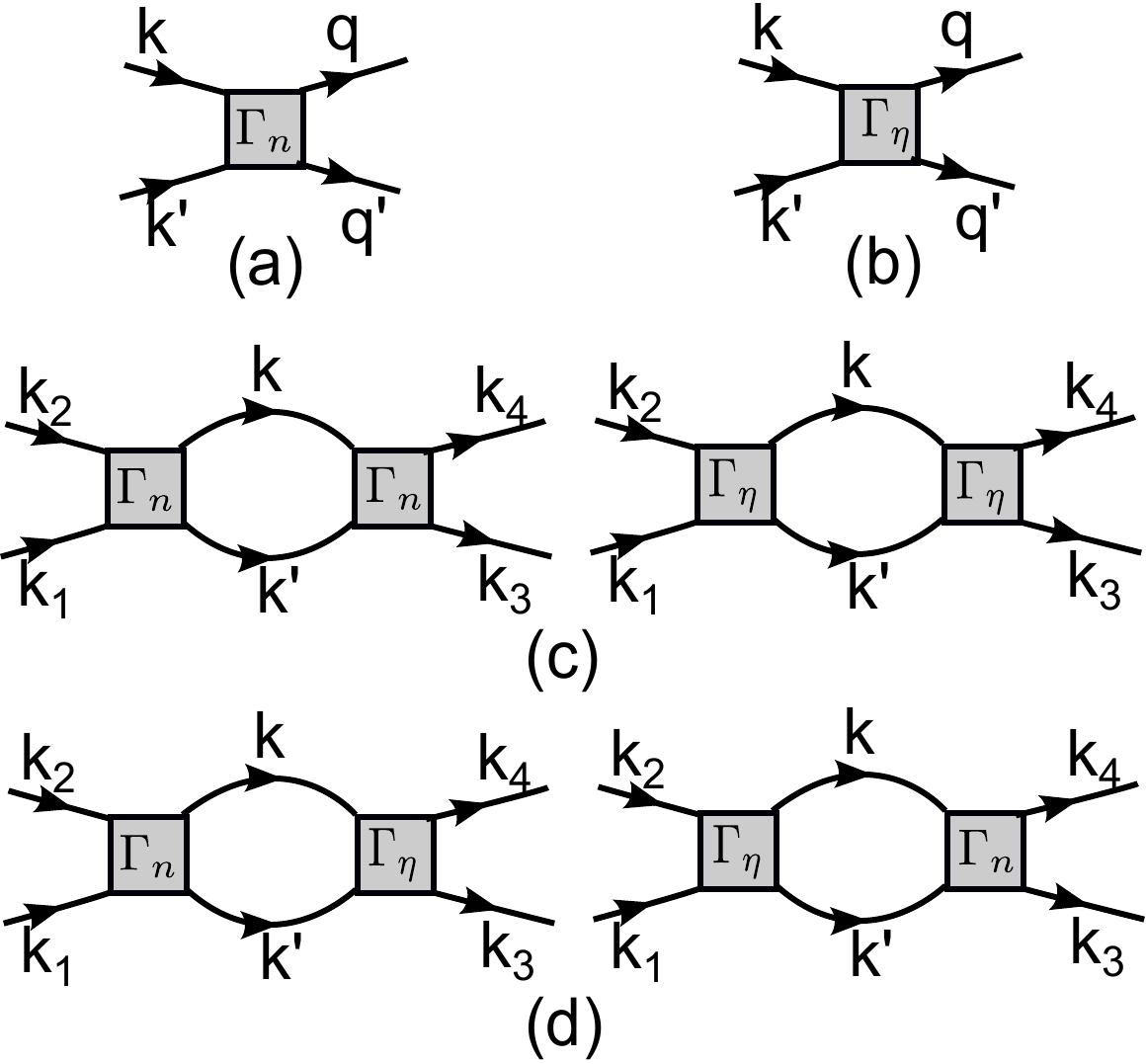}\\
  \caption{(a) The vertex corresponding to $\Gamma_{n}(\mathbf{k},\mathbf{k'},\mathbf{q},\mathbf{q'})$. (b) The vertex corresponding to $\Gamma_{\eta}(\mathbf{k},\mathbf{k'},\mathbf{q},\mathbf{q'})$. (c) The contributions to $V_n(\mathbf{k}_{1},\mathbf{k}_{2},\mathbf{k}_{3},\mathbf{k}_{4})$. (d) The contributions to $V_{\eta}(\mathbf{k}_{1},\mathbf{k}_{2},\mathbf{k}_{3},\mathbf{k}_{4})$. The four one-loop graphs in (c)(d) belong to particle-particle channel in FRG after introduce $\eta$ vertex $\Gamma_{\eta}$. We set all external momenta to Fermi momenta and the internal loop momenta are restricted to lie at the cutoff.}\label{feynmanvertex}
\end{figure}

\paragraph{Functional Renormalization Group}
FRG is an unbiased method for weak to moderate electron correlation systems. It calculates the different possible instabilities on an equal footing. And in each iteration of FRG, all virtual one-loop scattering processes including particle-particle, direct particle-hole and crossed particle-hole channels(Fig.~\ref{frgchannel}) are integrated out. In the previous study of FRG, the momenta of four-point vertex satisfy the momentum  conservation. In our letter, we make one important modification by introducing a $\eta$ vertex(Fig.~\ref{feynmanvertex}(a)), $\Gamma_{\eta}(\mathbf{k},\mathbf{k'},\mathbf{q},\mathbf{q'})$, which satisfies $\mathbf{k}+\mathbf{k'}=\mathbf{q}+\mathbf{q'}+\mathbf{Q}$, $\mathbf{Q}=(\pi,\pi)$. Otherwise we call the usual vertex normal vertex(Fig.~\ref{feynmanvertex}(b)), $\Gamma_{n}(\mathbf{k},\mathbf{k'},\mathbf{q},\mathbf{q'})$, which satisfies $\mathbf{k}+\mathbf{k'}=\mathbf{q}+\mathbf{q'}$. Consequently, each one-loop graph has four forms (taking particle-particle channel as examples, Fig.~\ref{feynmanvertex}(c,d)). Then FRG generates two type effective interactions: $V_n(\mathbf{k}_{1},\mathbf{k}_{2},\mathbf{k}_{3},\mathbf{k}_{4})$, $\mathbf{k}_{1}+\mathbf{k}_{2}=\mathbf{k}_{3}+\mathbf{k}_{4}$ and $V_{\eta}(\mathbf{k}_{1},\mathbf{k}_{2},\mathbf{k}_{3},\mathbf{k}_{4})$, $\mathbf{k}_{1}+\mathbf{k}_{2}=\mathbf{k}_{3}+\mathbf{k}_{4}+\mathbf{Q}$. Like in Ref.~\cite{Honerkamp2001,WangZhaiRanLee2009}, we also make some approximations in our calculation: ignoring the selfenergy-corrections on the internal lines of the one-loop corrections, ignoring the frequency dependence of the vertex function and projecting the external momenta onto the Fermi surface.
For the momentum space discretization we take the $N$-patch method\cite{Zanchi2000}, which divides the Brillouin zone into $N$ patches containing Fermi surface segments, thus the particle momentum dependence is treated by the patch index and in each patch the coupling function is approximated as a constant. More FRG technical details can be found in \cite{Metzner2012review,Honerkamp2010,ZhaiHui2009,Thomale2011}.

\begin{figure}
  \centering
  \includegraphics[width=9cm]{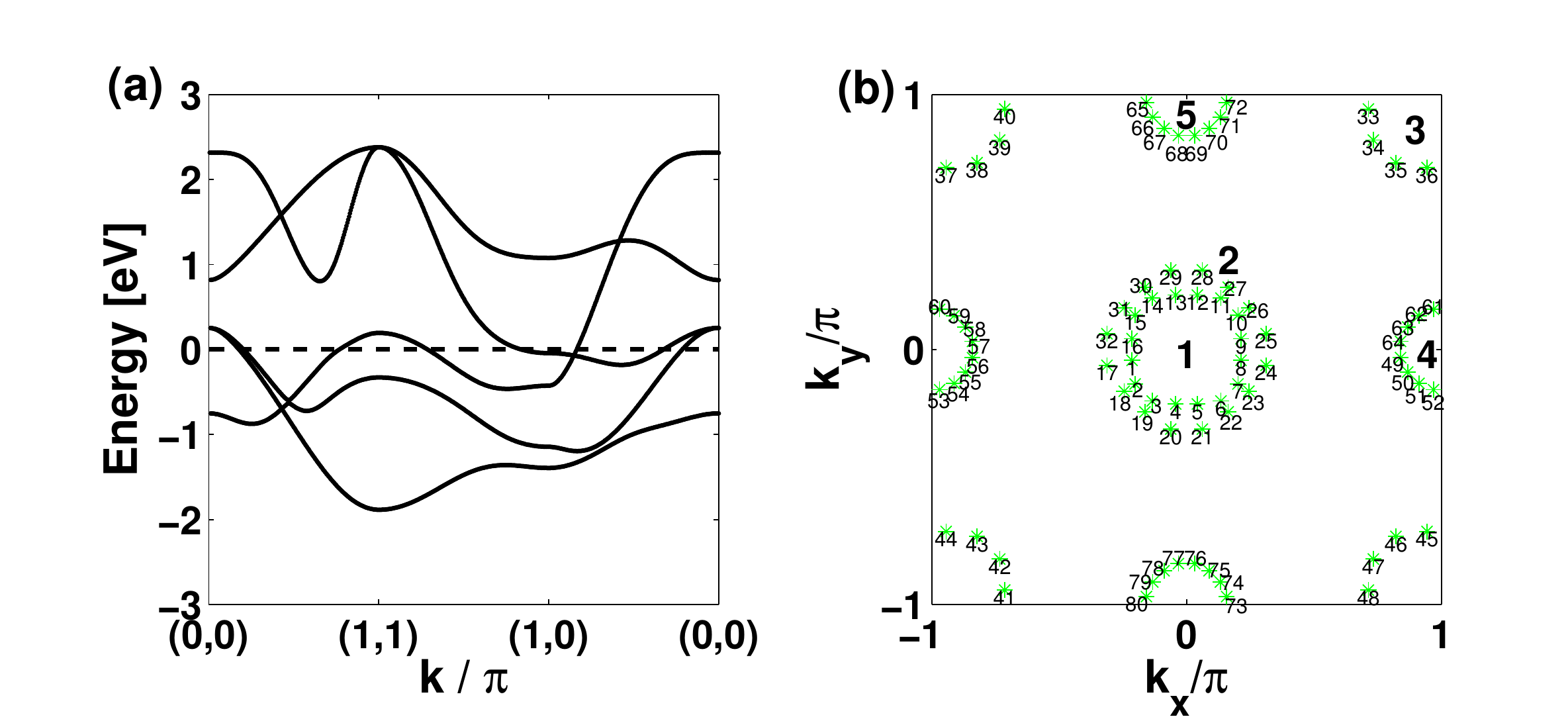}\\
  \caption{(a)(b) Band structure and Fermi surface used in this paper.  The tight binding model and parameters are given in \cite{Graser2010}; here the hole doping is 0.317. In (b) the Brillouin zone is divided into $N=80$ patches and the discrete green stars are the intersection point of patch center lines and Fermi surfaces. The number $1-5$ in plot (b) is the Fermi surface label, $FS_1$: small hole pocket at $(0,0)$, $FS_2$: large hole pocket at $(0,0)$, $FS_3$: hole pocket at $(\pi,\pi)$, $FS_4$: electron pocket at $(\pi,0)$ and $FS_5$: electron pocket at $(0,\pi)$. }\label{band_patch}
\end{figure}

\paragraph{Model}
We adopt the  five-orbital tight-binding model developed by Graser et al.\cite{Graser2010} which fits the full density-function-theory band structure for  iron-pnictides:
\begin{equation}
H_0 = \sum_{\mathbf{k},\sigma}\sum_{a,b=1}^{5} ( \xi_{ab}(\mathbf{k}) + \epsilon_{a}\delta_{ab} )c_{a\sigma}^{\dagger}(\mathbf{k})c_{b\sigma}(\mathbf{k})
\label{H0}\end{equation}
where $a,b$ stand for the five $d$ orbitals, $\sigma$ stands for the spin, $\xi_{ab}(\mathbf{k})$ is the kinetic term, $\epsilon$ is the on-site energy, $c_{a\sigma}^{\dagger}(\mathbf{k})$ creates a electron with spin $\sigma$ and momentum $\mathbf{k}$ in orbital $a$. The parameters used in Eq.~\ref{H0} can be found in Ref.~\cite{Graser2010} and in the later calculation we adjust some hopping parameters to tune the geometry shape of Fermi surfaces and then analysis the different Fermi surface nesting cases. The band structure and Fermi surface at 0.317 hole doping are illustrated in Fig.~\ref{band_patch}(a) and (b), respectively. There are five disjoint Fermi surfaces, and for simplicity we label the Fermi surfaces with numbers $1-5$(see Fig.~\ref{band_patch}(b)).

For the interaction part, we follow the natural choice to introduce intra- and inter-orbital Coulomb repulsion $U$ and $U^\prime$, plus Hund's coupling $J_H$ and pairing hopping $J_{pair}$ terms:

\begin{eqnarray}
H_{int} & = & \sum_i[U\sum_a n_{ia\uparrow}n_{ia\downarrow} + U^\prime\sum_{\substack{a<b \\ \sigma,\sigma^\prime}}n_{ia\sigma}n_{ib\sigma^\prime}\nonumber \\
 &  & +\sum_{a<b} ( J_H \sum_{\sigma,\sigma^\prime}c_{ia\sigma}^{\dagger}c_{ib\sigma^\prime}^{\dagger}c_{ia\sigma^\prime}c_{ib\sigma} \nonumber \\
 & & +J_{pair}c_{ia\uparrow}^{\dagger}c_{ia\downarrow}^{\dagger}c_{ib\downarrow}c_{ib\uparrow} )]
\end{eqnarray}

where $i$ labels the sites of a square lattice, $\sigma,\sigma^\prime$ label the spin, and $n_{ia\sigma}$ is number operators at site $i$ of spin $\sigma$ in orbital $a$. In the rest of the paper we set the intra-orbital coupling as $U=4.0eV$\cite{WangZhaiRanLee2009,Haule2008,WangZhaiLee2009}, and the other parameters are $U^\prime=2.0eV, J_H=J_{pair}=0.3eV$ so that it approximately satisfies $U=U'+2J_H$ \cite{WangZhaiRanLee2009,Fresard1997}. We begin the FRG calculation with the bare Hamiltonian $H_0+H_{int}$.

\begin{figure}
  \centering
  \includegraphics[width=9cm]{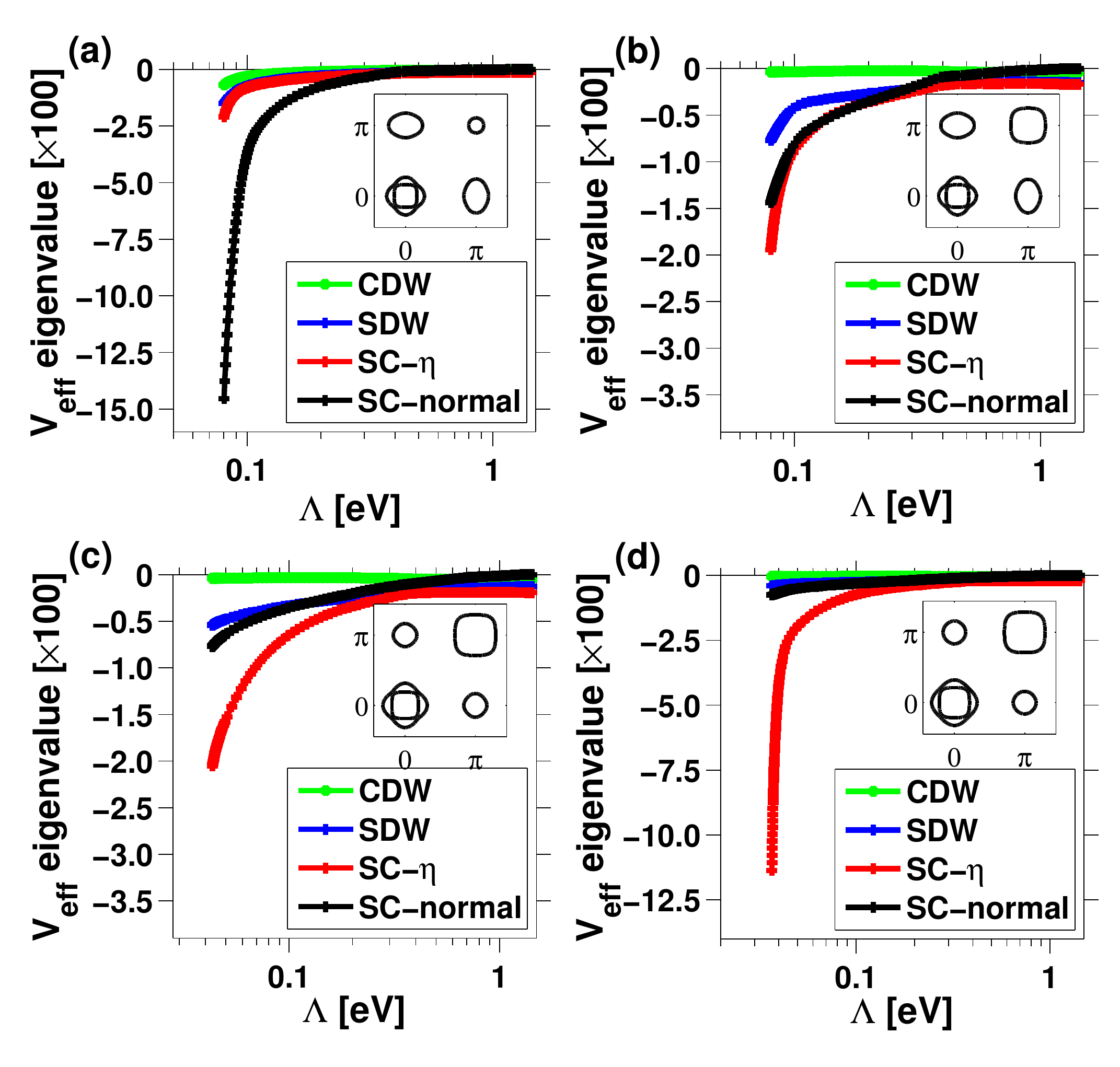}\\
  \caption{The FRG flows of CDW, SDW, SC $\eta$ pairing and SC normal pairing instabilities for different Fermi surface nesting cases, respectively. The corresponding Fermi surfaces are plotted in the insets. (a) Case (1), doping is 0.072 hole doping. (b) Case (2), 0.156 hole doping. (c) Case (3), 0.286 hole doping. (d) Case (4), 0.317 hole doping. The detailed information for these four cases is introduced in our letter.}\label{flow_kz0}
\end{figure}

We take the $N=80$ patches Brillouin zone discretization method for the five-Fermi-surface   and the setup of the patches around Fermi surface is shown in Fig.~\ref{band_patch}(b). In order to substantiate our perspective effectively, we tune the shape of Fermi surfaces upon varying some of the hopping parameters in the tight binding model(\cite{Graser2010}).  $Case (1): $ the parameters different from Ref.~\cite{Graser2010} are $t_x^{11}=-0.0704$, $t_x^{44}=0.2265$, $t_y^{11}=-0.3205$, $t_{xx}^{11}=0.0303$, $t_{xy}^{11}=0.2448$, $t_{xxy}^{11}=-0.0454$, $t_{xyy}^{11}=-0.0267$. In this case, the FRG flows of charge-density-wave (CDW), spin-density-wave (SDW), superconducting (SC) normal pairing and SC $\eta$ pairing instabilities for 0.072 hole doped system are shown in Fig.~\ref{flow_kz0}(a) and the corresponding Fermi surface is illustrated in the inset plot of Fig.~\ref{flow_kz0}(a). The most prominent instability channel is the SC normal pairing. This situation is representative for weak $(\pi,\pi)$ nesting between Fermi surfaces $\alpha$ and $\beta$ ($\alpha$, $\beta$ are Fermi surface labels defined in Fig.~\ref{band_patch}(b), ($\alpha,\beta$)=(1,3),(2,3),(4,5)). $Case (2):$ the flows of CDW, SDW, SC normal pairing and SC $\eta$ pairing instabilities for $t_x^{11}=-0.0704$, $t_y^{11}=-0.3205$, $t_{xx}^{11}=0.0303$, $t_{xy}^{11}=0.2448$, $t_{xxy}^{11}=-0.0454$, $t_{xyy}^{11}=-0.0267$ at 0.156 hole hoping are shown in Fig.~\ref{flow_kz0}(b). As energy cutoff $\Lambda$ decreasing, the SC $\eta$ pairing channel and SC normal pairing become competitive and $\eta$ pairing channel surpassing normal pairing channel becomes the leading instability at the minimum of $\Lambda$. In the inset of Fig.~\ref{flow_kz0}(b) we present the Fermi surface in this situation, we note that the nesting between $FS_2$ and $FS_3$ is slightly stronger compared with the previous case. $Case (3):$ the FRG flows of these instabilities for $t_x^{11}=-0.0704$, $t_y^{11}=-0.3205$, $t_{xx}^{11}=0.0303$, $t_{xy}^{11}=0.2448$, $t_{xxy}^{11}=-0.0454$, $t_{xyy}^{11}=-0.0267$, at 0.286 hole doping is shown in Fig.~\ref{flow_kz0}(c). The strength of SC $\eta$ pairing instability continue to enhance with perfect nesting between two electron pockets. $Case (4):$ in this case, we adopt the hopping parameters in Ref.~\cite{Graser2010} and set the chemical potential as $-0.20$. The SC $\eta$ pairing channel has a strong divergence while the other instabilities remain far behind in the FRG flows(Fig.~\ref{flow_kz0}(d)). From the Fermi surface shapes(Fig.~\ref{flow_kz0}(d)inset), we see it clearly the $(\pi,\pi)$ nesting is the strongest among the four cases.

\begin{figure}
  \centering
  \includegraphics[width=9.2cm]{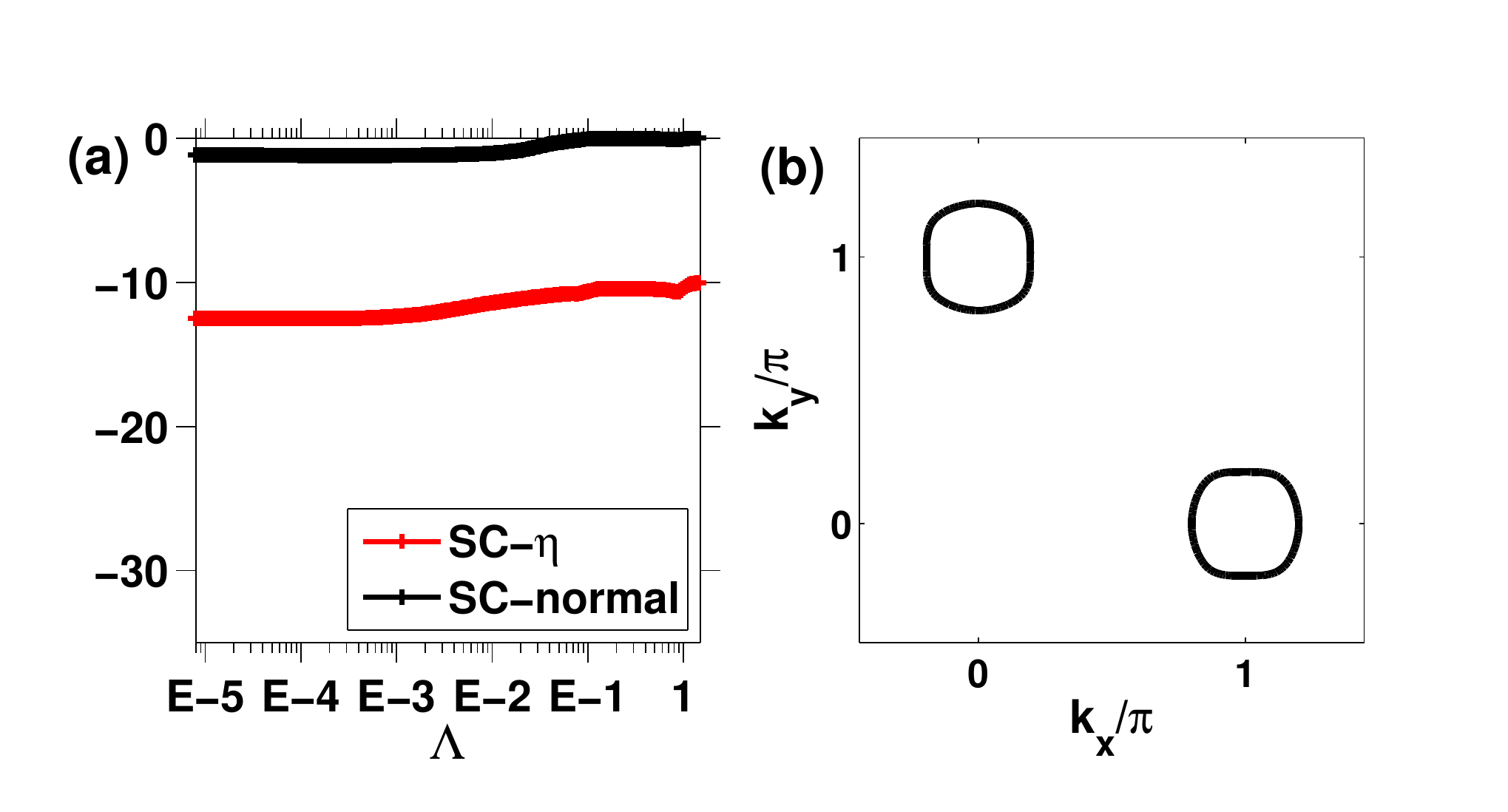}\\
  \caption{(a) The FRG flows of SC $\eta$ pairing and SC normal pairing instabilities for the case with only two electron pockets. (b) The two electron Fermi pockets.}\label{epocketonly}
\end{figure}

 We  can also analyze the case only electron pockets surviving, a case for 122 Iron-Chalcogenides\cite{Dagotto2013-review} and the single-layer FeSe\cite{Zhang2014-fese,He2013-fese,Liu2012-fese,Tan2013-fese}. The hole bands are suppressed and sink below the Fermi level through tuning the hopping parameters as $t_{x}^{11}=-0.1604$, $t_y^{11}=-0.4005$ $t_z^{44}=0.0201$, $t_{xz}^{44}=0.0722$, $t_{xyz}^{44}=0.0391$, $t_{xyz}^{35}=-0.0004$. In this condition, there are only two electron pockets centered at $(\pi,0)$ and $(0,\pi)$ while all the hole pockets disappear(Fig.~\ref{epocketonly}(b)). The SC $\eta$ pairing and  normal pairing channels are both very weak.  They have no dispersion tendency as the energy cutoff $\Lambda$ decreasing, furthermore, $\eta$ pairing is always stronger than normal pairing channel, as shown in Fig.~\ref{epocketonly}(a). It is clear that FRG, which is essentially a weak coupling approach,  is not well suited to address the high $T_c$ in this situation. Nevertheless, this calculation suggests that the $\eta$ pairing between electron pockets  can be very important.

\begin{figure}
  \centering
  \includegraphics[width=6cm]{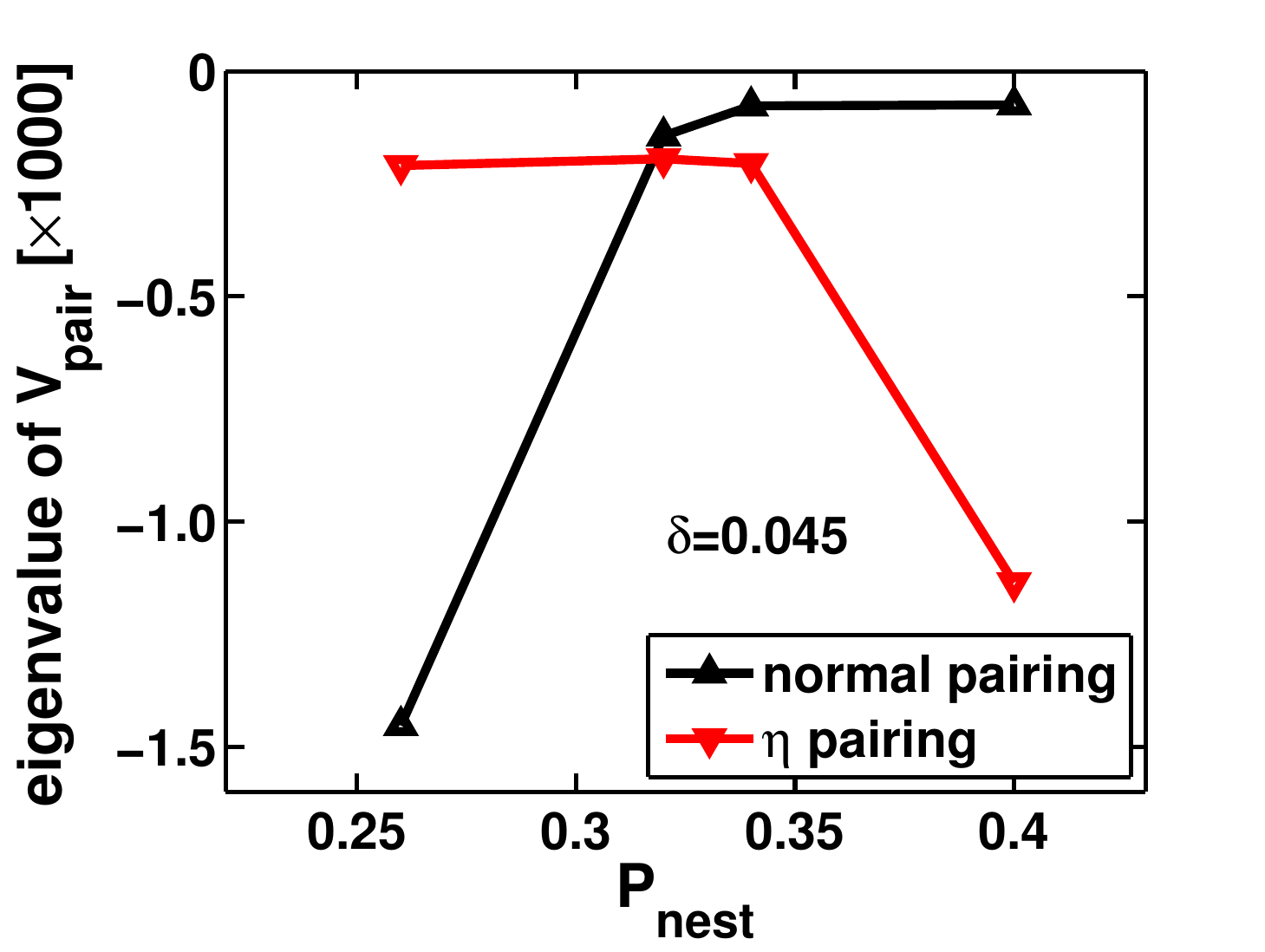}\\
  \caption{The leading eigenvalues of SC normal pairing and SC $\eta$ pairing versus nesting scale $P_{nest}$ at the minimum of the energy cutoff $\Lambda$. At low nesting scale SC normal pairing is the leading pairing order, while at high nesting scale SC $\eta$ pairing becomes the leading channel. }\label{pnest}
\end{figure}

Through comparative analysis of the above results, we find that the divergence rate of SC $\eta$ pairing instabilities are closely related to the nesting condition linked by a $(\pi,\pi)$ nesting vector. We define a Nesting Scale $P_{nest}$ to measure the degree of $(\pi,\pi)$ Fermi surface nesting.

\begin{eqnarray}
P_{nest}=\frac{1}{5N}\sum_{(\alpha,\beta)}\sum_{k=1}^{N}\theta(\delta-|p_{f\alpha}(k)-p_{f\beta}(k^{\prime})|)
\end{eqnarray}
and $\theta(\delta-|p_{f\alpha}(k)-p_{f\beta}(k^{\prime})|)$ is a step function:
\begin{eqnarray}
\theta(\delta -|p_{f\alpha} (k)- p_{f\beta} (k')|) = \left\{ \begin{array} {cc} 1 & |p_{f\alpha}(k)- p_{f\beta}(k')| \le \delta \\ 0 & |p _{f\alpha}(k)- p_{f\beta}(k')|> \delta \end{array} \right.
\end{eqnarray}

where $\alpha$ and $\beta$ are Fermi surface indices, and ($\alpha,\beta$)=(1,3), (2,3), (4,5), for the nesting vector is $(\pi,\pi)$. $k$ is patch index, $k'$ associates with $k$. $p_f(k)$ is the length of Fermi vector with particles locating at the $k$-patch. The value of $\delta$ is about one-tenth of Fermi surface radius, here we take  $\delta = 0.045$.
We calculate the nesting scale of the four cases in Fig.~\ref{flow_kz0}.  Then we extract the eigenvalues of SC normal pairing and SC $\eta$ pairing at the minimum of the energy cutoff $\Lambda$ from FRG results. The relation between $P_{nest}$ and the strength of two types of superconducting instabilities is shown in Fig.~\ref{pnest}. This result proves that the $\eta$ pairing is determined by the nesting condition between two hole pockets.  The normal pairing strength which has been intensively studied before is known to be strongly dependent on the nesting condition between electron and hole pockets linked by the nesting wavevector $(0,\pi)$ or $(\pi,0)$\cite{WangZhaiLee2009,Thomale2011}. Moreover, it is known in the 1-Fe BZ that the hole pockets at $(0,0)$  are mainly attributed to $d_{xz,yz}$ orbitals and the hole pocket at $(\pi,\pi)$ is dominated by $d_{xy}$ orbital. Therefore, the $\eta$ pairing suggested by FRG is  mainly an inter-orbital $\eta$ pairing. We notice that when we are finishing this paper.

While FRG is valid in the weak coupling limit,  it is fashion to  seek an effective model at low energy in strong coupling limit that can be responsible for the results. Previously, the multi-orbital  $t-J_1-J_2$ model has been suggested\cite{Seo2008,Si2008}. With only the normal pairing,  it has shown that only the intra-orbital   next-nearest-neighbor (NNN) AFM   coupling $J_2$  causes a dominant normal pairing in the s-wave channel. In this FRG result, it is clear that the inter-orbital  AFM term is responsible for the $\eta$ pairing. Therefore, as the $\eta$ and normal pairings can be both spontaneously formed, a nearest-neighbor  inter-orbital AFM coupling, $J'_1$, should be added to the previous model. With such a model, if we take the phenomenological arguments recently promoted in Ref.~\cite{Huding2012,Davis2014}, the dominant $\eta$ pairing could be in the p-wave spin singlet channel.  In this case, the mixed state with an s-wave normal pairing and a p-wave $\eta$ pairing  has even parity. This state does not  break parity conservation but could break time-reversal symmetry.   A study of such a model will be reported elsewhere. When we are finishing our paper,  we notice that the state in a new paper\cite{Lin2014-ku} may be consistent with our results here.

In summary,
with the FRG approach, we have shown that the $\eta$ pairing is an important instability what can take place in iron-based superconductors. In this weak coupling method, the $\eta$ pairing is driven by the  $(\pi,\pi)$ Fermi surface nesting between two hole pockets with different orbital characters, which suggests that the $\eta$ pairing is an inter-orbital pairing.  We suggest that the effective model at low energy  should include  the nearest-neighbour  inter-orbital AFM exchange coupling to account for the $\eta$ pairing.

We thank BY Liu, R. Thomale and F Wang
for useful discussion and help on FRG method. The work is supported by the Ministry of Science and Technology of China 973
program(2012CB821400) and NSFC.


\begin{thebibliography}{10}

\bibitem{Hirschfeld2011}
P.~J. Hirschfeld, M.~M. Korshunov, and I.~I. Mazin, Rep. Prof. Phys. {\bf 74},
  124508  (2011).

\bibitem{Johnston2010-review}
D.~C. Johnston, Adv. Phys. {\bf 59},  803  (2010).

\bibitem{Dagotto2013-review}
E. Dagotto, Rev. Mod. Phys. {\bf 85},  849  (2013).

\bibitem{BCS1957}
J. Bardeen, L.~N. Cooper, and J.~R. Schrieffer, Phys. Rev. {\bf 108},
  1175  (1957).

\bibitem{Yang1989}
C.~N. Yang, Phys. Rev. Lett. {\bf 63},  2144  (1989).

\bibitem{Hu2013}
J. Hu, Phys. Rev. X {\bf 3},  031004  (2013).

\bibitem{Khodas2012}
M. Khodas and A.~V. Chubukov, Phys. Rev. Lett. {\bf 108},  247003  (2012).

\bibitem{Mazin2009}
I.~I. Mazin and J. Schmalian, Physica C {\bf 469},  614
  (2009).

\bibitem{Yigao2010}
Y. Gao and S. Wu-Pei, Phys. Rev. B {\bf 81},  104504  (2010).

\bibitem{Hu2014-odd}
N. Hao and J. Hu, Phys. Rev. B {\bf 89},  045144  (2014).

\bibitem{Hu2012s4}
J. Hu and N. Hao, Phys. Rev. X {\bf 2},  021009  (2012).

\bibitem{Tan2013-fese}
S.~Y. Tan {\it et~al.}, Nat. Mater. {\bf 12},  634  (2013).

\bibitem{Zhang2013-impurity}
P. Zhang {\it et~al.}, Arxiv:1312.7064  (2013).

\bibitem{Shankar1994}
R. Shankar, Rev. Mod. Phys. {\bf 66},  129  (1994).

\bibitem{Metzner2012review}
W. Metzner {\it et~al.}, Rev. Mod. Phys. {\bf 84}, (2012).

\bibitem{Honerkamp2010}
C. Honerkamp,  Euro. Phys. J. Spec. Top. {\bf 188},  33
  (2010).

\bibitem{Honerkamp2001}
C. Honerkamp, M. Salmhofer, N. Furukawa, and T. Rice, Phys. Rev. B {\bf 63},
  18  (2001).

\bibitem{WangZhaiLee2009}
F. Wang, H. Zhai, and D.-H. Lee, Europhys. Lett. {\bf 85},  37005  (2009).

\bibitem{Thomale2011}
R. Thomale {\it et~al.}, Phys. Rev. Lett. {\bf 107},  117001  (2011).

\bibitem{Graser2010}
S. Graser {\it et~al.}, Phys. Rev. B {\bf 81},  214503  (2010).

\bibitem{Seo2008}
K. Seo, B.~A. Bernevig, and J. Hu, Phys. Rev. Lett. {\bf 101},  206404
  (2008).

\bibitem{Si2008}
Q. Si and E. Abrahams, Phys. Rev. Lett. {\bf 101},  76401  (2008).

\bibitem{WangZhaiRanLee2009}
F. Wang {\it et~al.}, Phys. Rev. Lett. {\bf 102},  047005  (2009).

\bibitem{Zanchi2000}
D. Zanchi and H.~J. Schulz, Phys. Rev. B {\bf 61},  13609  (2000).

\bibitem{ZhaiHui2009}
H. Zhai, F. Wang, and D.-H. Lee, Phys. Rev. B {\bf 80},  064517  (2009).

\bibitem{Haule2008}
K. Haule, J.~H. Shim, and G. Kotliar, Phys. Rev. Lett. {\bf 100},  226402
  (2008).

\bibitem{Fresard1997}
R. Fr\'eard and G. Kotliar, Phys. Rev. B {\bf 56},  12909  (1997).

\bibitem{Zhang2014-fese}
Q.~E. Wang and F.~C. Zhang, arXiv  1401.7159  (2014).

\bibitem{He2013-fese}
S.~L. He {\it et~al.}, Nat. Mater. {\bf 12},  605  (2013).

\bibitem{Liu2012-fese}
D. Liu {\it et~al.}, Nat. Commun. {\bf 3},  931  (2012).

\bibitem{Huding2012}
J. Hu and H. Ding, Sci. Rep. {\bf 2},  381  (2012).

\bibitem{Davis2014}
J.~S. Davis and D.-H. Lee, Proc. Nat.
Acad. Sciences {\bf 110},  17623  (2013).

\bibitem{Lin2014-ku}
C.-H. Lin, C.-P. Chou, W.-G. Yin, and W. Ku, arXiv:1403.3687  (2014).

\end{thebibliography}

%
%

\end{document}